# Abnormal Respiratory Sound Identification Using Audio-Spectrogram Vision Transformer

Whenty Ariyanti, Kai-Chun Liu, Kuan-Yu Chen, Yu Tsao

*Abstract—* Respiratory disease, the third leading cause of deaths globally, is considered a high-priority ailment requiring significant research on identification and treatment. Stethoscope-recorded lung sounds and artificial intelligence-powered devices have been used to identify lung disorders and aid specialists in making accurate diagnoses. In this study, audio-spectrogram vision transformer (AS-ViT), a new approach for identifying abnormal respiration sounds, was developed. The sounds of the lungs are converted into visual representations called spectrograms using a technique called short-time Fourier transform (STFT). These images are then analyzed using a model called vision transformer to identify different types of respiratory sounds. The classification was carried out using the ICBHI 2017 database, which includes various types of lung sounds with different frequencies, noise levels, and backgrounds. The proposed AS-ViT method was evaluated using three metrics and achieved 79.1% and 59.8% for 60:40 split ratio and 86.4% and 69.3% for 80:20 split ratio in terms of unweighted average recall and overall scores respectively for respiratory sound detection, surpassing previous state-of-the-art results.

## I. INTRODUCTION

According to the World Health Organization (WHO), chronic respiratory disorders (CRDs) are among the leading causes of death globally [1]. Chronic obstructive pulmonary disease (COPD) alone is responsible for about 6% of deaths globally and is a progressive, fatal lung disease that impairs airflow in the lungs, leading to increased risks of exacerbations. Although it cannot be cured, treatment can help reduce the risk of mortality and alleviate symptoms. Chronic pulmonary disease is not a single illness but encompasses a group of disorders affecting lung function, such as asthma, COPD, occupational lung diseases, and pulmonary hypertension.

Early detection of this respiratory disease is crucial, and pulmonary auscultation has been a standard part of medical examinations since the 19th century [2]. Pulmonary auscultation is a non-invasive, quick, inexpensive, and simple method that can be performed well even by inexperienced physicians or laypeople. However, the success of this method depends largely on the examiner's experience and ear acuity. The lungs have several auscultation points in the chest, sides, and back, with different sound characteristics corresponding to different lung sections and chest anatomy. To accurately diagnose, the stethoscope must be positioned properly on the lung surface to differentiate the lung sounds from other background noise.

Recent technological advancements have led to the development of digital stethoscopes that can record lung sounds and save them on computers. The integration of digital signal processing techniques with lung sound analysis has become possible in recent years and has been used to improve diagnostic efficiency. This research was conducted to develop an approach that yields more consistent and reliable results with greater efficiency and can be supervised, if necessary.Advancements in wireless technologies and IoT have extended the reach of automated diagnoses, making them more widely available and accessible when combined with cloud services. However, the complexity of lung physiology leads to highly variable sound dynamics that are influenced by factors such as location, patient posture, airflow intensity, age, weight, and gender, making signal processing in lung sound analysis a challenging task [3]. Consequently, experts may offer different subjective interpretations of the same sounds. Therefore, establishing a standardized set of characteristics that could serve as indicators for specific illnesses is difficult, which hinders the implementation of automatic diagnoses.

In this paper, we proposed and evaluated a novel audio-spectrogram vision transformer (AS-ViT) model to detect and categorize abnormal lung sounds. The AS-ViT involves converting the spectrogram of the analyzed waveform into patches, flattening and embedding them into a sequence, and then applying position embedding to preserve position information. The resulting sequence was processed through multiple attention layers to generate a final representation, which was then fed into a SoftMax classification layer for classification. The objective of this work was to provide a reliable tool for healthcare professionals to assess lung diseases associated with these sounds, with the aim of improving the accuracy and robustness of anomalous sound detection and classification in noisy or challenging environments.

The remainder of this paper is organized as follows. The proposed AS-ViT system is described in Section 2. The experiments and results are discussed in Section 3. Finally, the conclusions and future work are presented in Section 4.

## II. THE PROPOSED AS-VIT SYSTEM

In this section, we present AS-ViT, a model approach for generating generic audio representations.

W. Ariyanti, and K.-Y. Chen are with the Department of Computer Science and Information Engineering, National Taiwan University of Science and Technology, Taipei 10607, Taiwan.

W. Ariyanti, K.-C. Liu, and Y. Tsao are with the Research Center for Information Technology Innovation (CITI) at Academia Sinica, Taipei 115201, Taiwan.

(Corresponding author: W. Ariyanti, and Y. Tsao, e-mail: ariyantiwhenty@gmail.com, and yu.tsao@citi.sinica.edu.tw).

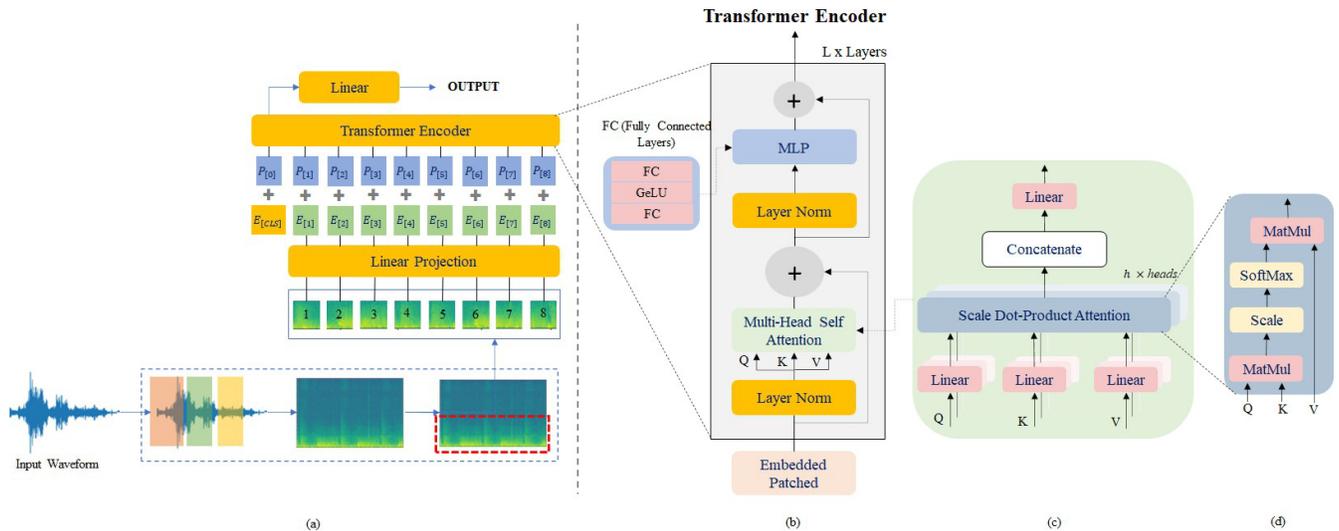

Fig. 1. The proposed Audio Spectrogram Vision Transformer (AS-ViT) architecture: (a) the primary architecture of the proposed model; (b) the Transformer encoder module; (c) the Multiscale-self attention (MSA) head, and (d) the self-attention (SA) head.

*A. Audio-Spectrogram Vision Transformers (AS-ViT)*

The audio waveform of t seconds is transformed into a spectrogram, as illustrated in Figure 1(a) 1 (a). A log-mel spectrogram was used instead of the raw waveform as the input for the AS-ViT model. Spectrograms are widely used in the audio, speech, and music fields because they contain abundant low-level acoustic information, which resembles images, making them a suitable choice for ViTs. Let $S = \{X_i, y_i\}_{i=1,...n}^{r}$, represent a set of spectrogram images. Here $X_i$, is an spectrogram image and $y_i$ is its corresponding label $y_i \in \{1, 2, 3, .., m\}$, where $m$ is the number of defined class for the set.

The ViT architecture is based entirely on the traditional Transformer design [4], The design has received considerable attention in recent years owing to its exceptional performance in tasks such as machine translation and other natural language processing (NLP) applications. At the highest level, an image is split into smaller segments called patches, and a series of linear representations of these patches are used as input for a transformer. Patches in an image are processed similarly to tokens (words) in NLP applications. The Transformer uses an encoder-decoder architecture and can process sequential input without the need for a recurrent network, owing to its self-attention mechanism, which enables it to capture long-range connections between different parts of the sequence. This mechanism has contributed significantly to the success of transformer models.

Transformers underperform when trained on small datasets, because they lack the inductive biases found in CNNs, such as the ability to focus on specific areas of an image. However, when trained on large datasets, they have been shown to achieve or exceed the current state-of-the-art performance on many image recognition benchmarks. The architecture of the model is shown in Figure 1. It consists of a transformer encoder that takes spectrogram images as input. In the first stage, Picture $X$ from the training set is divided into non-overlapping patches. Each patch is treated as a separate token by the transformer before being fed into the encoder. Two separate classifiers, the token, and distiller classifiers are attached to the top of the encoder. During the testing phase, the average of both classifiers is considered the final prediction. The model components are discussed in detail subsequently.

*1) Encoder Module:* The data-efficient image transformer architecture, a modified version of the ViT model that requires less training data, is used in the encoder part of the model. The spectrogram is transformed into a sequence of N 16 × 16 patches, with an overlap of six in both the time and frequency dimensions, before being input into the model. N is the number of patches and the effective input sequence length for the transformer. It is calculated as N = 12d (100t - 16)/10e.

*2) Linear Projection Layer:* Before being fed into the encoder, a sequence of patches is transformed into a vector with a model dimension of d through linear projection using a learned embedding matrix E. This vector, along with a learnable classification token, is concatenated to perform classification. The positional information of the patches is also added to the patch representations to maintain their original spatial arrangements.

*3) Vision Transformer Encoder:* The sequences of the embedded patches are transmitted to the transformer encoder, as illustrated in Figure 1(b). The transformer encoder comprises $L$ identical layers, each composed of two key elements: (1) a multi-head self-attention block (MSA), and (2) a fully-connected feed-forward dense block multilayer perceptron (MLP). The MLP block contains two dense layers separated by a GeLU activation function. Both components of the encoder layer have residual skip connections and are normalized using layer normalization. The final output of the encoder is the first element in the sequence, $z_L^0$, which is then sent to an external classifier for class label prediction.

$$y = LN(z_L^0) \quad (1)$$

The MSA block in the Transformer encoder is a critical part of the transformer encoder. Its role is to assess the importance of a given patch embedding relative to other embeddings in the sequence. As illustrated in Figure 1(c), the MSA block consists of four layers: a linear layer, self-attention layer, concatenation layer that combines the outputs of multiple attention heads, and final linear layer, which evaluates the relative importance of each patch embedding in the sequence by computing the weighted sum of all Z values. The SA head is responsible for computing the dot product among the query, key, and value vectors to calculate the attention weights for each sequence patch. The SA block is composed of four layers: a linear layer, self-attention layer, concatenation layer (which merges the outputs from multiple attention heads), and final linear layer. Fig. 1(d) shows the computation within the SA block. The input sequence is multiplied by three learned matrices, $U_{QKV}$ query ($Q$), key ($K$), and value ($V$) to produce three values for each element. To determine the importance of each sequence patch, the SA block calculates the dot product between the query vector of an element and the key vectors of all other elements. The dot-product results are then scaled and fed into a softmax function. The SA block performs the dot-product operation similar to a regular dot product, but includes the $D_k$ dimension as a scaling factor. The softmax output is multiplied by the value vector of each embedded patch to determine the patch with the highest attention score. This entire process can be summarized by three equations.

$$[Q, K, V] = zU_{QKV}, U_{QKV} \in \mathbb{R}^{d \times 3D_K} \quad (2)$$

$$A = softmax(\frac{QK^T}{\sqrt{D_K}}), A \in \mathbb{R}^{n \times n} \quad (3)$$

$$SA(z) = A.V \quad (4)$$

The MSA block in the encoder calculates the attention scores for each patch embedded in the sequence, relative to the others. This is achieved by performing the scaled dot-product attention calculation h times, using different Q, K, and V matrices each time. The outputs from each of the h heads are then combined and transformed using a feed-forward layer with learnable weights $w$. The mathematical expression for the operation is as follows:

$$MSA(z) = Concat(SA_1(z); SA_2(z); ...SA_h(z))W, W \in \mathbb{R}^{h \cdot D_k x} \quad (5)$$

## III. EXPERIMENTS

In this section, we describe the experimental setup, including the data preparation process. We then present the results of our experiments on AS-ViT and the analysis of our findings.

### A. Data Acquisition

The 2017 ICBHI dataset includes 920 recordings from 126 patients spanning a total of 5.5 h. Each respiration cycle in the recordings has been classified by an expert as normal, crackle, wheeze, or both crackle and wheezing. The duration of these cycles varies, as shown in Figure 2, ranging from 0.2 s to 16.2 s (mean cycle length: 2.7 s) across the four classes.

The audio recordings are obtained using four distinct types of instruments: the AKGC417L Microphone, 3M Littmann Classic II SE Stethoscope, 3M Littmann 3200 Electronic Stethoscope, and the WelchAllyn Meditron Master Elite Electronic Stethoscope. Previous research revealed that uneven representation of patients and respiration cycles across different classes results in a distorted data distribution, which can affect model performance significantly.

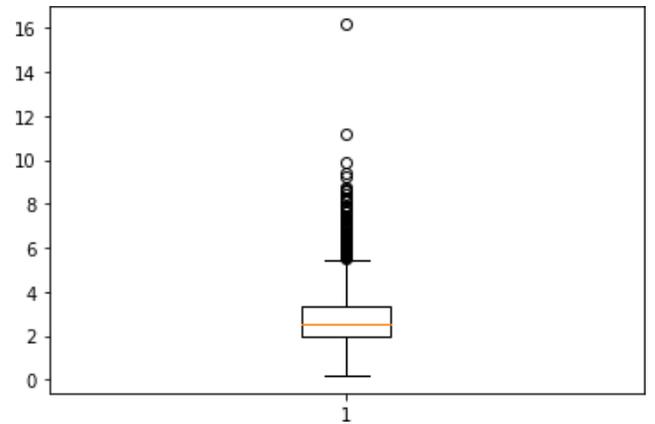

Fig. 2. Respiratory Cycles Duration

### B. Experimental Setup

Before training the model, further data preparation steps were performed. First, information about crackles and wheezing, including the start and end times in seconds, was obtained by reading each item on the list. A list was created to gather this data, which included the patient ID and type of recording (stereo or mono) for later use. Additionally, the audio file was divided into smaller parts to isolate the breathing section of the sound, as indicated by the start and end times listed in the text files.

To keep all spectrograms the same size, regardless of their different lengths, a duration of 10 s was set for each instance. Instances longer than 10 s were trimmed, and shorter instances were filled with zeros. During the training phase, the instances were randomly trimmed, During the test phase, however, they were trimmed from a predetermined starting point to maintain consistency in the evaluation methods. Additionally, to enhance the accuracy of the outcomes, a subject-independent division was implemented on the initial training dataset. Specifically, samples for training and validation were split 60:40 and 80:20, respectively. The original test set was used for evaluation.

Throughout our experiments, we employed the ViT-Base model, which was established using the parameters outlined in [5]. The model comprised 12 encoder layers, each of which featured 12 attention heads, an embedding dimension of 768, and a feed-forward subnetwork size of 3072. The model was initially pre-trained on the Imagenet-21k dataset, which included 14 million images and 21,843 classes, and subsequently fine-tuned on the Imagenet-1k dataset.

*C. Evaluation Metrics*

In this study, for ease of comparison with other models, the proposed model was evaluated using the same measurement as that used in the 2017 ICBHI assessment. The evaluation metrics were the average score (AS), which is the average sensitivity (SE) and specificity (S), as stated in the challenge [6].

$$Sensitivity, S_e = \frac{P_c + P_w + P_b}{Crackle, Wheeze, Both} \quad (6)$$

$$Specificity, S_p = \frac{P_n}{Normal} \quad (7)$$

$$Score = \frac{S_e + S_p}{2} \quad (8)$$

where $P_c$, $P_w$, $P_b$, and $P_n$ represent the number of accurate predictions in the four categories of lung sounds: crackle, wheeze, both, and normal sounds, respectively.

*D. Experimental Results*

Finally, the results of the proposed method were compared with the state-of-the-art results in the literature. The performance of the proposed AS-ViT approach for respiratory sound detection was compared with those of previously reported systems. A comparison of the performance of extant models and other approaches using the same dataset and settings is presented in Table I. To categorize respiratory sounds into four different categories, we adhere to the challenge evaluation metrics and utilize the official 60:40 data split. The results indicated that the AS-ViT model outperformed the other models in terms of recall and score, with a recall of 79.13% and a score of 59.84% for a 60:40 split ratio and achieved 86.4% and 69.3% for a 80:20 split ratio, respectively. However, in terms of precision, the ARSCNet model from [10] performed better with a precision of 46.38%. This indicates that the proposed model can identify and classify instances more accurately than other models.

## IV. CONCLUSION

In this study, a new approach was developed for identifying abnormal respiratory sounds using AS-ViT. The proposed method was tested using the ICBHI 2017 challenge dataset, and it outperformed current state-of-the-art methods in terms of classification accuracy. Additionally, it was demonstrated that the application of vision transformer techniques to respiratory data could further enhance the performance evaluation and achieved 79.1% and 59.8% for 60:40 split ratio and 86.4% and 69.3% for 80:20 split ratio in terms of unweighted average recall and overall scores respectively for respiratory sound detection, surpassing previous state-of-the-art results. Future studies are recommended to investigate alternative techniques for transformer compression and create lightweight models.

TABLE I
EXPERIMENTAL RESULTS FROM ICBHI 2017 RESPIRATORY DATA

| Model | Precision | Recall | Score |
|---|---|---|---|
| **Split Ratio 60:40** | | | |
| LungBRN [7] | 31.1 | 69.2 | 50.2 |
| LungRN+NL [8] | 41.3 | 63.2 | 52.3 |
| CNN+CBA+BRC [9] | 40.2 | 71.8 | 55.3 |
| CNN+CBA+BRC+FT [9] | 40.1 | 72.3 | 59.8 |
| ARSCNet [10] | **46.4** | 67.1 | 56.8 |
| **AS-ViT (ours)** | 40.6 | **79.1** | **59.8** |
| **Split Ratio (80:20)** | | | |
| LungRN+NL [8] | 63.7 | 64.7 | 64.2 |
| CNN+CBA+BRC [9] | 54.4 | 79.7 | 67.1 |
| CNN+CBA+BRC+FT [9] | 53.7 | 83.3 | 68.5 |
| **AS-ViT (ours)** | 52.1 | **86.4** | **69.3** |